\documentclass[twocolumn,aps,prl,reprint]{revtex4-1}
\usepackage[latin9]{inputenc}
\usepackage{amsmath}
\usepackage{amssymb}
\usepackage{graphicx}

\makeatletter
\@ifundefined{textcolor}{}
{%
 \definecolor{BLACK}{gray}{0}
 \definecolor{WHITE}{gray}{1}
 \definecolor{RED}{rgb}{1,0,0}
 \definecolor{GREEN}{rgb}{0,1,0}
 \definecolor{BLUE}{rgb}{0,0,1}
 \definecolor{CYAN}{cmyk}{1,0,0,0}
 \definecolor{MAGENTA}{cmyk}{0,1,0,0}
 \definecolor{YELLOW}{cmyk}{0,0,1,0}
}

\usepackage{color}

\@ifundefined{textcolor}{}{%
 \definecolor{BLACK}{gray}{0}
 \definecolor{WHITE}{gray}{1}
 \definecolor{RED}{rgb}{1,0,0}
 \definecolor{GREEN}{rgb}{0,1,0}
 \definecolor{BLUE}{rgb}{0,0,1}
 \definecolor{CYAN}{cmyk}{1,0,0,0}
 \definecolor{MAGENTA}{cmyk}{0,1,0,0}
 \definecolor{YELLOW}{cmyk}{0,0,1,0}
}

\def\NOT(#1,#2){\OneQubitGate(#1,#2){$X$}}

\makeatother

\begin{document}

\title{Towards the speed limit of high fidelity 2-qubit gates}

\author{Swathi S. Hegde}\email{swathi.hegde@tu-dortmund.de}
\altaffiliation[Current address: ]{Quantum Brilliance GmbH 5.0G, 
Industriestrasse 4, 70565 Stuttgart, 
Germany.}
\author{Jingfu Zhang}\email{jingfu.zhang@tu-dortmund.de}
\author{Dieter Suter}\email{Dieter.Suter@tu-dortmund.de}
\affiliation{ Fakult\"{a}t Physik, Technische Universit\"{a}t Dortmund,\\
 D-44221 Dortmund, Germany}
\begin{abstract}
Most implementations of quantum gate operations rely on external control
fields to drive the evolution of the quantum system. Generating these
control fields requires significant efforts to design the suitable
control Hamiltonians. Furthermore, any error in the control fields
reduces the fidelity of the implemented control operation with respect
to the ideal target operation. Achieving sufficiently fast gate operations
at low error rates remains therefore a huge challenge. In this work,
we present a novel approach to overcome this challenge by eliminating,
for specific gate operations, the time-dependent control fields entirely.
This approach appears useful for maximising the speed of the gate
operation while simultaneously eliminating relevant sources of errors.
We present an experimental demonstration of the concept in a single
nitrogen-vacancy (NV) center in diamond at room temperature. 
\end{abstract}
\maketitle
\textit{Introduction}.--Quantum gates are the elementary steps for
processing quantum information. They are therefore essential for all
quantum technologies, such as quantum computing \cite{shor1999polynomial,grover1996fast,georgescu2014quantum}
or quantum sensing \cite{degen2008scanning,taylor2008high,degen2017quantum}
and they must be implemented in all types of quantum registers such
as superconducting qubit systems \cite{devoret2013superconducting,chow2012universal},
ion traps \cite{kielpinski2002architecture,cirac1995quantum}, or
hybrid qubit systems, combining, e.g., electronic and nuclear spins
\cite{degen2008scanning,taylor2008high,degen2017quantum}. In most
cases, elementary quantum gates are realized by segments of external
control fields, often including free evolution periods \cite{nielsen,stolze,daskin2011decomposition,divincenzo1995two,barenco1995conditional}.
Designing these sequences of control fields is an optimisation task,
where the number of control field segments, their strengths, durations
and phases are adjusted such that the resulting unitary has maximum
overlap with the target quantum gate \cite{hegde2019efficient,taminiau2012detection,khaneja2005optimal,lovchinsky2016nuclear}.
The duration of the gates is limited by the strength of the couplings
between the qubits and by the strength of the interaction between
the qubits and the control fields \cite{3093,oliveira2011nmr,pati2002quantum}.

These limitations become severe when gate operation times exceed the
qubit coherence times and thus coherent control becomes impossible.
Although various techniques were demonstrated for alleviating this
problem like protected quantum gates \cite{zhang2014protected,van2012decoherence}
or indirect control \cite{zhangic,taminiau2012detection}, they add
control overhead and the resulting gate durations still tend to be
long. Here, we introduce a novel approach to overcome these challenges,
which results in highly efficient gates that have the shortest possible
durations without any control overhead.

The computational basis states of a single qubit are $|0\rangle$
and $|1\rangle$, and for a 2-qubit system $\{|00\rangle,|01\rangle,|10\rangle,|11\rangle\}$.
While the computational basis states are usually assigned to eigenstates
of the Hamiltonian, we choose here a different approach where some
computational states are not eigenstates of the system Hamiltonian.
This allows us to generate logical operations such as a conditional
rotation (CR) without applying control fields, simply by allowing
the system to evolve under its internal Hamiltonian. In the example
discussed below, the states $|00\rangle$ and $|01\rangle$ are eigenstates
of the system Hamiltonian and hence do not evolve, while $|10\rangle$
and $|11\rangle$ are superpositions of eigenstates. The evolution
of this system therefore generates a conditional gate operation with
the first qubit acting as the control qubit. The evolution is periodic,
with period $t_{p}=2\pi/|{\cal E}_{3}-{\cal E}_{4}|$, where ${\cal E}_{i}$
are the energies of the eigenstates and we choose units where $\hbar=1$.
Thus for a delay $\tau$, the free evolution implements a conditional
rotation $U_{CR}(\alpha)$ with the rotation angle $\alpha=2\pi\tau/t_{p}=\tau|{\cal E}_{3}-{\cal E}_{4}|$,
such that $U_{CR}(\alpha)$ reaches the quantum speed limit given
by the energy of the system \cite{3093,oliveira2011nmr,stolze}. The
resulting $U_{CR}(\alpha)$ is the fastest possible gate for the given
Hamiltonian and is not affected by errors in control fields. In this
paper, we provide details on how this type of gate operations can
be implemented and show experimental results.

\begin{figure}[b]
\centering \includegraphics[width=8.5cm]{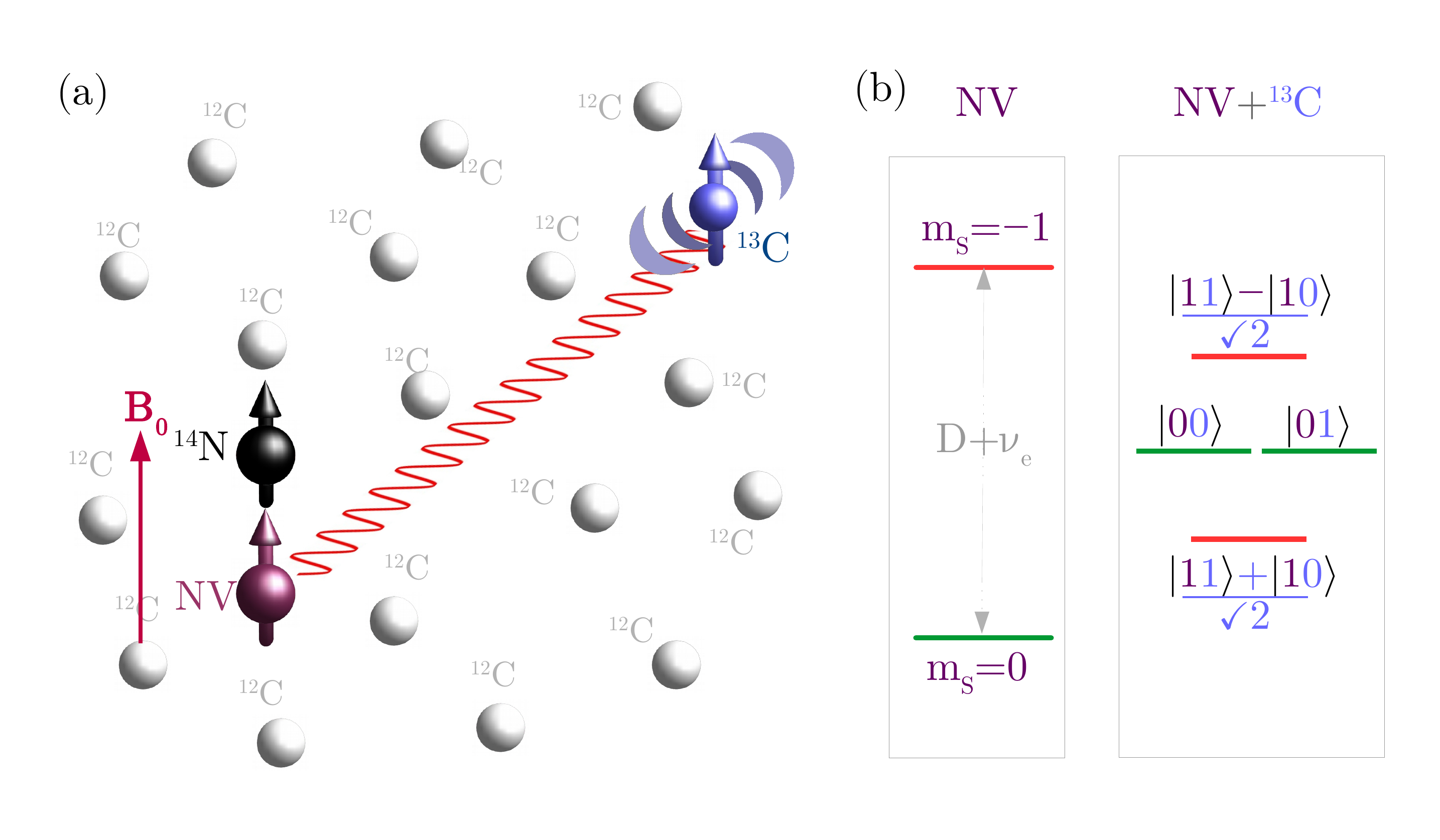} \caption{(a) Structure of the nitrogen-vacancy (NV) system consisting of an
electron and two nuclear spins: $^{14}$N and $^{13}$C. The magnetic
field is oriented along the NV axis which we take to be the $z-$axis.
The $^{13}$C atom is at a distance of $0.89$ nm from the electron
and hence is weakly coupled. The other nuclei are spinless $^{12}$C.
(b) Energy levels of the electron (left) and system subspace consisting
of two electron spin states and the $^{13}$C.}
\label{nv} 
\end{figure}

\textit{Hamiltonian and computational basis.--} For the demonstration
of this fast gate operation, we use a single nitrogen-vacancy (NV)
center in diamond at room temperature \cite{gaebel2006room,neumann2008multipartite,maurer2013room,childress2006coherent,suter2017single}.
The NV center consists of a spin-1 electron coupled to a spin-1 $^{14}$N
and a spin-1/2 $^{13}$C, as shown in Fig. \ref{nv}(a). For the 2-qubit
operation, we choose a subsystem consisting of two of the electron
spin levels as the control qubit and the $^{13}$C spin as the target
qubit, while the $^{14}$N is a passive spin that is not affected
by the gate operations. The secular part of the electron-$^{13}$C
system Hamiltonian in the laboratory frame is

\begin{eqnarray}
\frac{{\cal {H}}}{2\pi} & = & DS_{z}^{2}-\nu_{e}S_{z}\nonumber \\
 &  & -\nu_{C}I_{z}+A_{zz}S_{z}I_{z}+A_{zx}S_{z}I_{x},\label{eq:Hf-1}
\end{eqnarray}
where $S_{z}$ and $I_{z/x}$ are the spin operators for the electron
and the $^{13}$C, respectively, $D=2.870$ GHz is the zero field
splitting of the electron, $A_{zz}=-0.152$ MHz and $A_{zx}=0.110$
MHz are the longitudinal and transverse components of the hyperfine
coupling with $^{13}$C, $\nu_{e}=(\gamma_{e}B_{0}-2.16)\mathrm{\,MHz}=-400.110$
MHz is the electron Larmor frequency that includes the shift from
the $^{14}$N hyperfine coupling when the nitrogen is in the $m_{N}=1$
state, and $\nu_{C}=\gamma_{C}B_{0}=0.152$ MHz is the $^{13}$C Larmor
frequency in a magnetic field $B_{0}=14.2$ mT. Here $\gamma_{e}$
and $\gamma_{C}$ are the gyromagnetic ratios of electron and $^{13}$C,
respectively.

In the presence of $B_{0}$, the two transitions between the electron
spin states $m_{S}=$$0\leftrightarrow-1$ and $0\leftrightarrow+1$
are well separated with a frequency difference of $2\nu_{e}$. We
choose to implement our gates in the electron subspace $m_{S}=\{0,-1\}$
and define $s_{z}$ as the pseudo-spin-1/2 operator for this subspace,
with eigenvalues $\pm1/2$. Each of the two electron spin levels splits
into two due to the coupling with $^{13}$C. The resulting four levels
form our system subspace and we call its Hamiltonian the subspace
Hamiltonian $\mathcal{H}_{s}$ {[}see supplementary material (SM){]}.
We choose the eigenstates of $s_{z}$ and $I_{z}$ as the computational
basis states $\{|0\rangle,|1\rangle\}\otimes\{|0\rangle,|1\rangle\}$.

We transform the subspace Hamiltonian to an interaction frame as

\begin{eqnarray}
{\cal H}_{I} & = & {\cal U}_{tr}(\tau)\mathcal{H}_{s}{\cal U}_{tr}^{\dagger}(\tau)-i{\cal U}_{tr}(\tau)\frac{d{\cal U}_{tr}^{\dagger}(\tau)}{d\tau}\nonumber \\
 & = & |1\rangle\langle1|\otimes[-2\pi A_{zx}I_{x}]\label{Hr-1-2}
\end{eqnarray}
where the interaction frame is defined by the unitary ${\cal U}_{tr}(\tau)=\exp(-i2\pi\tau[\nu_{C}|0\rangle\langle0|\otimes I_{z}+(D+\nu_{e})s_{z}\otimes I-\frac{{(D+\nu_{e})}}{2}I\otimes I])$,
where $I$ is the $2\times2$ identity operator {[}see SM{]}. The
energy eigenstates of ${\cal H}_{I}$ are $\{|00\rangle,|01\rangle,|1\psi\rangle,|1\phi\rangle\}$,
where $|\psi\rangle=\frac{|1\rangle+|0\rangle}{\sqrt{2}}$ and $|\phi\rangle=\frac{|1\rangle-|0\rangle}{\sqrt{2}}$,
with eigenvalues ${\cal E}_{1}={\cal E}_{2}=0$ and ${\cal E}_{3}=-{\cal E}_{4}=\pi A_{zx}$,
as indicated in Fig. \ref{nv}(b).

With this Hamiltonian, a free evolution of duration $\tau$ generates
the propagator 
\begin{equation}
\begin{split}U_{CR}(\alpha) & =\exp(-i\mathcal{H}_{I}\tau)=|0\rangle\langle0|\otimes I+|1\rangle\langle1|\otimes R_{x}(\alpha)\end{split}
,\label{eq:Us-1}
\end{equation}
where $R_{x}(\alpha)=\exp(i\alpha I_{x})$. This corresponds, as intended,
to a controlled rotation gate: if the control qubit (the electron)
is in state $|0\rangle$, the target qubit (the nuclear spin) does
not evolve; if the control qubit is in state $|1\rangle$, the target
qubit is rotated around the $x$-axis.

\begin{figure}[t]
\centering \includegraphics[width=8cm]{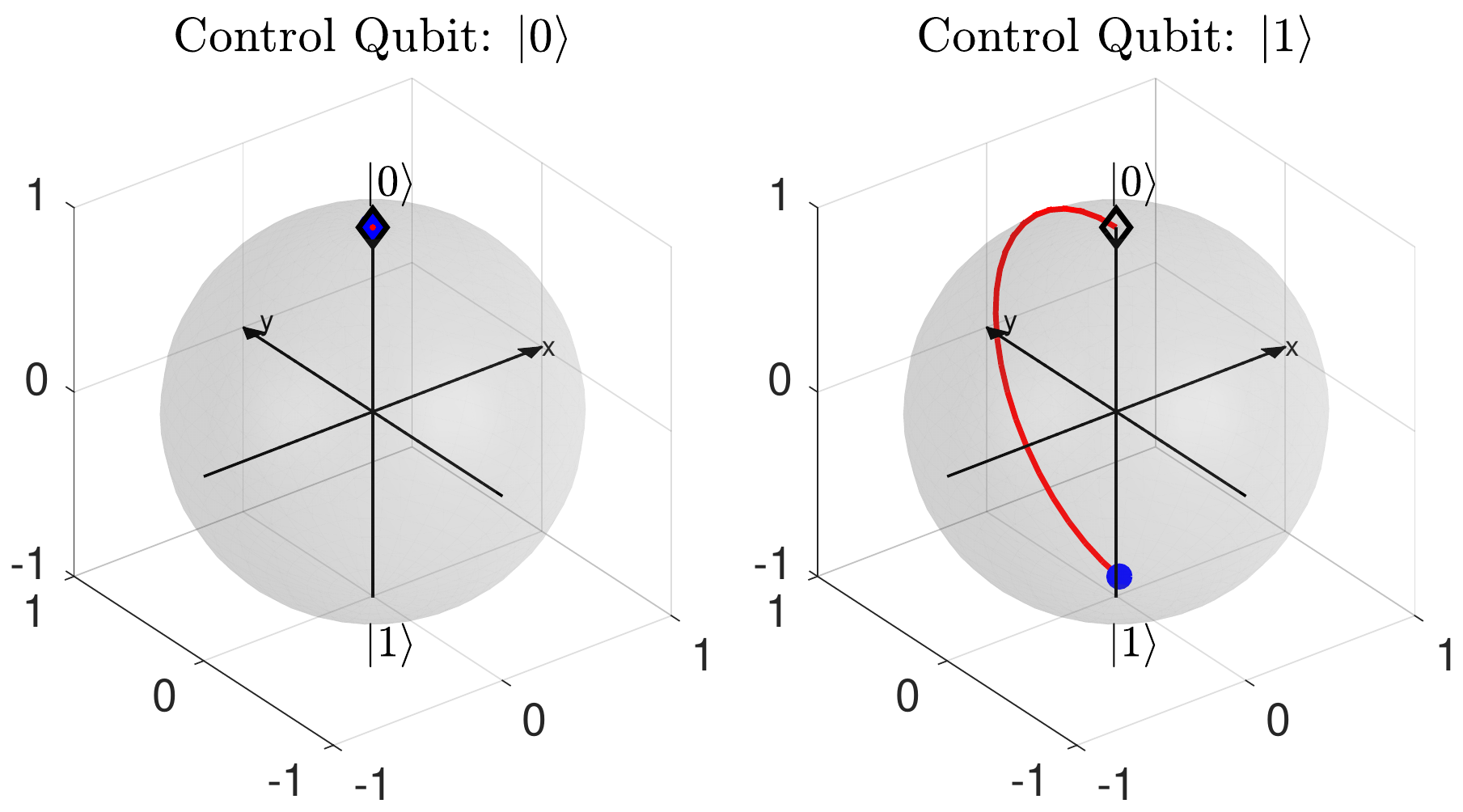} \caption{Bloch-sphere representation of the evolution of the $^{13}$C spin
initially in state $|0\rangle$, when the electron spin is in $|0\rangle$
(left) and $|1\rangle$ (right), during $U_{CR}(\pi)$. The black
diamonds and blue circles indicate the initial and the final states
respectively.}
\label{bloch} 
\end{figure}

The speed of a logical gate is limited by the average energy of the
quantum system \cite{3093,oliveira2011nmr,stolze}. The difference
between the energies of the eigenstates in the $m_{S}=-1$ subspace
is $2\pi A_{zx}$ and thus for the controlled rotation gate $U_{CR}(\alpha)$,
the minimum gate time is $\tau=\alpha/2\pi|A_{zx}|$. In our system,
where $A_{zx}=0.110$ MHz, a free evolution time of $\tau=4.545\,\mu$s
corresponds to $\alpha=\pi$ and reaches the quantum speed limit.
Fig. \ref{bloch} shows the evolution trajectory of the $^{13}$C
spin on the Bloch-sphere when it is intially in $|0\rangle$. If the
electron is in state $|0\rangle$, the nuclear spin does not evolve.
If it is in $|1\rangle$, the nuclear spin evolves from $|0\rangle$
$\rightarrow|1\rangle$ along a great circle trajectory, indicating
this is the fastest possible gate for the given system. 

\textit{Experimental demonstration.--} In our experiment we apply
the delay-only $U_{CR}(\alpha)$ to the states $|00\rangle$ and $|10\rangle$
to check its effect when the control qubit is in the state $|0\rangle$
and $|1\rangle$, respectively. After the $U_{CR}(\alpha)$ operation,
we measure the diagonal elements of the final density operator in
the computational basis \cite{zhang2021fast}.

The preparation of the initial state $|00\rangle$ from the maximally
mixed state of the spin system is as follows: a 532 nm wavelength
laser pulse of duration $5\,\mu$s and power $\approx0.5$ mW sets
the electron to its ground state $|0\rangle$ while the $^{13}$C
spin remains in the maximally mixed state. We then swap the states
of electron and $^{13}$C spins before resetting the electron spin
to the ground state using another laser pulse \cite{hegde2019efficient,zhangic}.
After the second laser pulse, the populations of the states $|00\rangle$
and $|01\rangle$ are 0.91 and 0.09, respectively \cite{zhangic,zhang2021fast}.
To further purify the state, we use a clean-up operation $U_{cu}$
of the form $(90_{y}^{\circ}-d-90_{x}^{\circ})$, which selectively
removes the spurious population of $|01\rangle$ from our system subspace
to $m_{S}=+1$ \cite{zhang2018pulse}. Here $90_{x/y}^{\circ}$ are
MW pulses acting on the electron spin, with the carrier frequency
set to the transition $m_{S}=0\leftrightarrow+1$, which rotate the
electron spin by $90^{\circ}$ around the $x/y-$axis and $d=1/2|A_{zz}|$
is the delay between the two $90_{x/y}^{\circ}$ pulses {[}see SM{]}.
As a result only $|00\rangle$ remains populated in the computational
subspace and the system subspace is fully polarised. A similar MW
pulse sequence with opposite pulse phases of the form $(90_{x}^{\circ}-d-90_{y}^{\circ})$
selectively removes the population of $|00\rangle$ from the system
subspace to $m_{S}=+1$ \cite{zhang2021fast,zhang2018pulse} and we
call this different clean-up operation $V_{cu}$ {[}see SM{]}. Below,
we show how $V_{cu}$ can be incorporated as a part of our readout
process.

In the NV system, readout of the state of the system is performed
by counting photons during a laser pulse. The resulting count rate
is a measure of the population of the ground state $m_{S}=0$; the
signal therefore represents the sum of the two populations $P_{|00\rangle}$
and $P_{|01\rangle}$ in the states $|00\rangle$ and $|01\rangle$,
respectively.

\begin{figure}[t]
\centering \includegraphics[width=8.5cm]{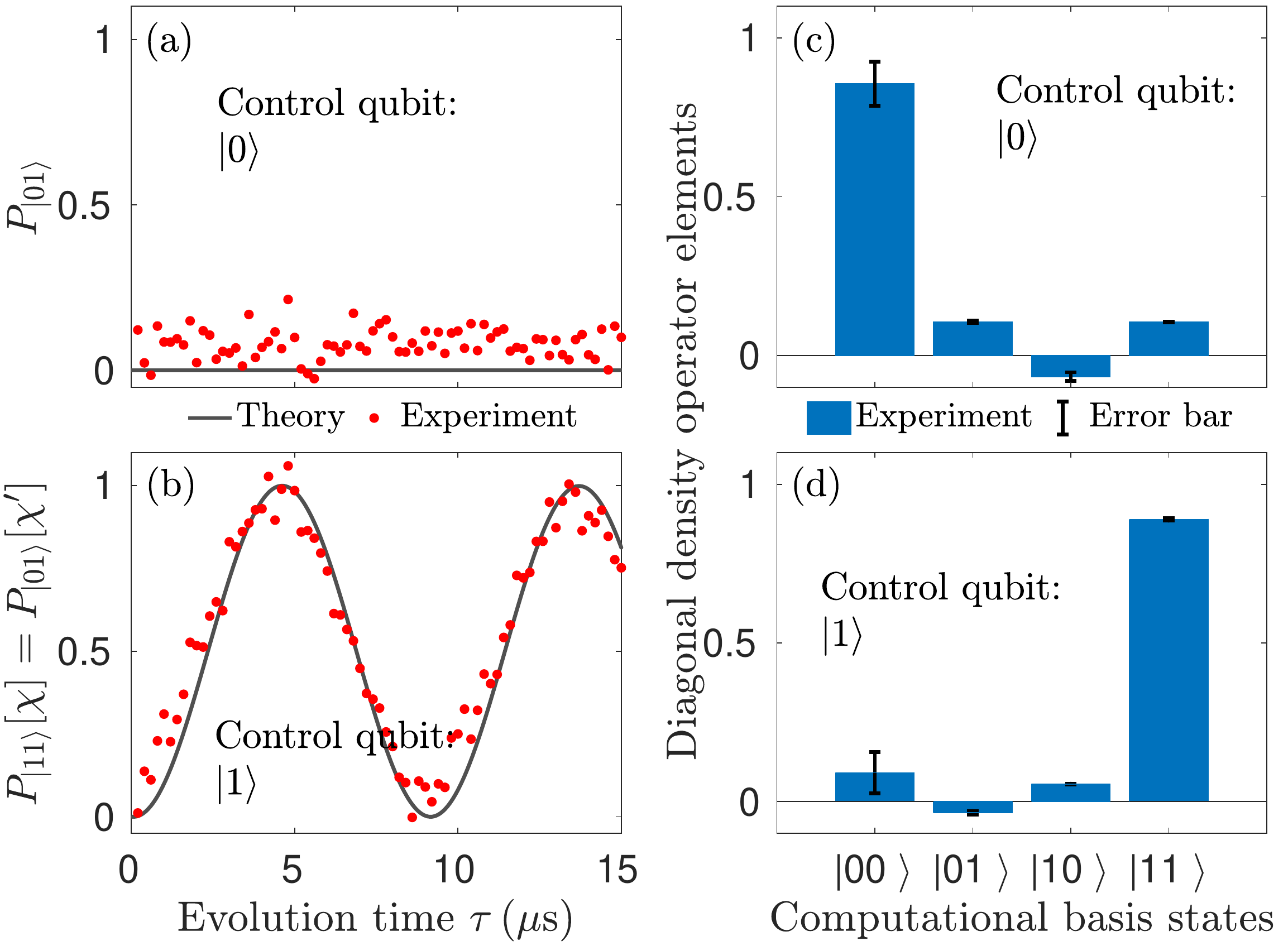}

\caption{Theoretical and experimental results to demonstrate the effect of
$U_{CR}(\alpha)$ using the delay-only sequence starting from the
initial states $|00\rangle$ (top) and $|10\rangle$ (bottom). (a,
b) The plots represent $P_{|01\rangle}$ (a) and $P_{|11\rangle}$
(b) as a function of the evolution time $\tau$. $P_{|11\rangle}[\chi]=P_{|01\rangle}[\chi']=[1-\cos(2\pi A_{zx}\tau)]/2$
oscillates from the initial value of 0 to a maximum of $1$ after
$\tau=4.545\,\mu$s, indicating the effect of $U_{CR}(\pi)$. (c,
d) Diagonal density operator elements of the final states in the computational
basis.}
\label{expt-1} 
\end{figure}

To check the effect of $U_{CR}(\alpha)$ when the control qubit is
in $|0\rangle$, we let the initial state $|00\rangle$ evolve under
${\cal H}_{I}$ for a variable duration $\tau=\alpha/2\pi|A_{zx}|$.
We then implement another clean-up operation $V_{cu}$ and thus the
population measured during the readout laser pulse corresponds to
$P_{|01\rangle}$. In Fig. \ref{expt-1}(a), we show the corresponding
theoretical and experimental plots of $P_{|01\rangle}$ vs $\tau$.

\begin{figure}[b]
\centering \includegraphics[width=7cm]{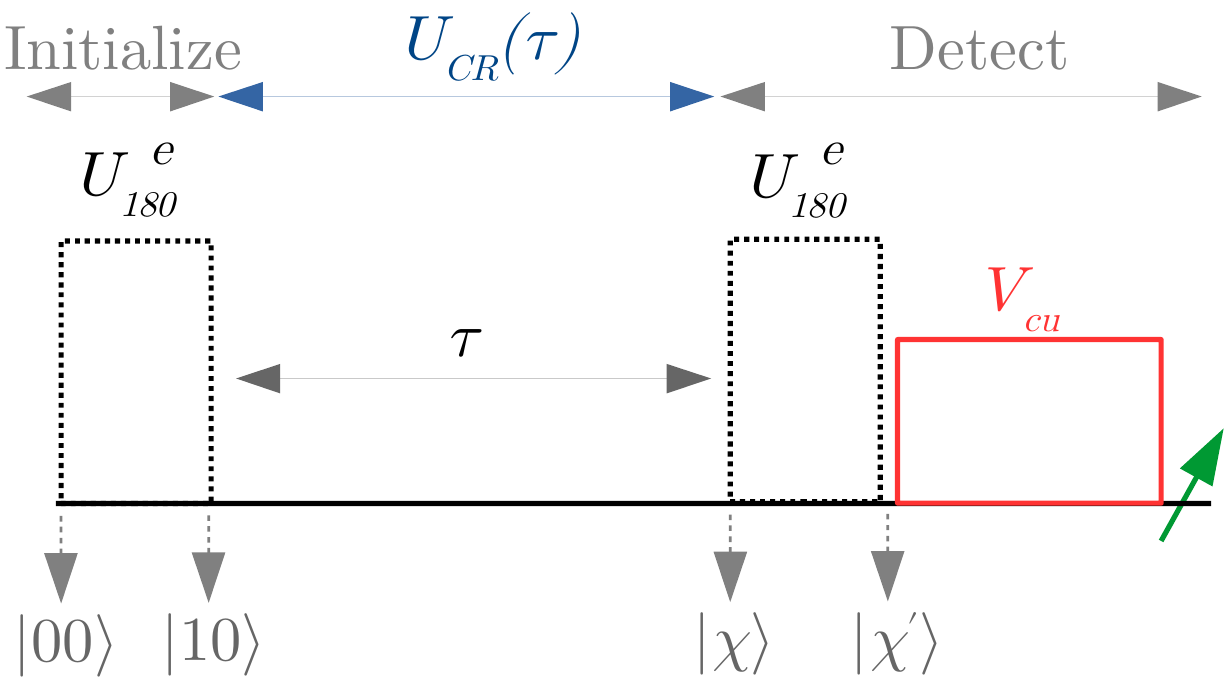}

\caption{Pulse sequence to check the effect of $U_{CR}(\alpha)$ when the control
qubit is in state $|1\rangle$.}
\label{ps} 
\end{figure}

We now check the effect of $U_{CR}(\alpha)$ when the control qubit
is in $|1\rangle$ by preparing the initial state $|10\rangle$, using
the pulse sequence given in Fig. \ref{ps}: starting from $|00\rangle$,
we apply an operation $U_{180}^{e}=\exp(-i\pi s_{x})$ as a MW pulse
resonant with the transition $m_{S}=0\leftrightarrow-1$ with an amplitude
of 7 MHz that rotates the electron spin by $180^{\circ}$ and exchanges
the states $|00\rangle$ $\leftrightarrow$ $|10\rangle$. During
the subsequent delay $\tau$, $|10\rangle$ evolves to $|\chi\rangle=\cos(\pi A_{zx}\tau)|10\rangle+i\sin(\pi A_{zx}\tau)|11\rangle$
under the Hamiltonian ${\cal H}_{I}$. The numerical simulation of
the probability $P_{|11\rangle}(\tau)=[1-\cos(2\pi A_{zx}\tau)]/2$
of finding the system in $|11\rangle$ vs $\tau$ is shown in Fig.
\ref{expt-1}(b).

In order to measure $P_{|11\rangle}$, we apply another $U_{180}^{e}$
operation to $|\chi\rangle$ to flip the electron states $m_{S}=-1$
$\leftrightarrow$ $m_{S}=0$ such that $|\chi'\rangle=U_{180}^{e}|\chi\rangle=\cos(\pi A_{zx}\tau)|00\rangle+i\sin(\pi A_{zx}\tau)|01\rangle$
followed by another clean-up operation $V_{cu}$. Since $U_{180}^{e}$
flips $|11\rangle$ $\leftrightarrow$ $|01\rangle$, the measured
signal represents $P_{|11\rangle}[\chi]=P_{|01\rangle}[\chi']$. The
experimental results of $P_{|11\rangle}$ vs $\tau$ is shown in Fig.
\ref{expt-1}(b).

We also measured the other diagonal elements of the final density
operators using the procedure of Ref. \cite{zhang2021fast}. The results
shown in Figs. \ref{expt-1}(c, d) verify that the state remains unchanged
when the control qubit is in $|0\rangle$ and the target qubit is
flipped when the control qubit is in $|1\rangle$. The experimental
state fidelities are $>0.98$ and $>0.99$ for Figs. \ref{expt-1}(c)
and (d) respectively. The state fidelity is $F=\frac{\mathrm{Tr}(\rho\rho_{t})}{\sqrt{\mathrm{Tr}(\rho^{2})\mathrm{Tr}(\rho_{t}^{2})}}$,
where $\rho_{t}$ is the target state and $\rho$ is the experimental
final state.

\begin{figure}[t]
\centering \includegraphics[width=8.5cm]{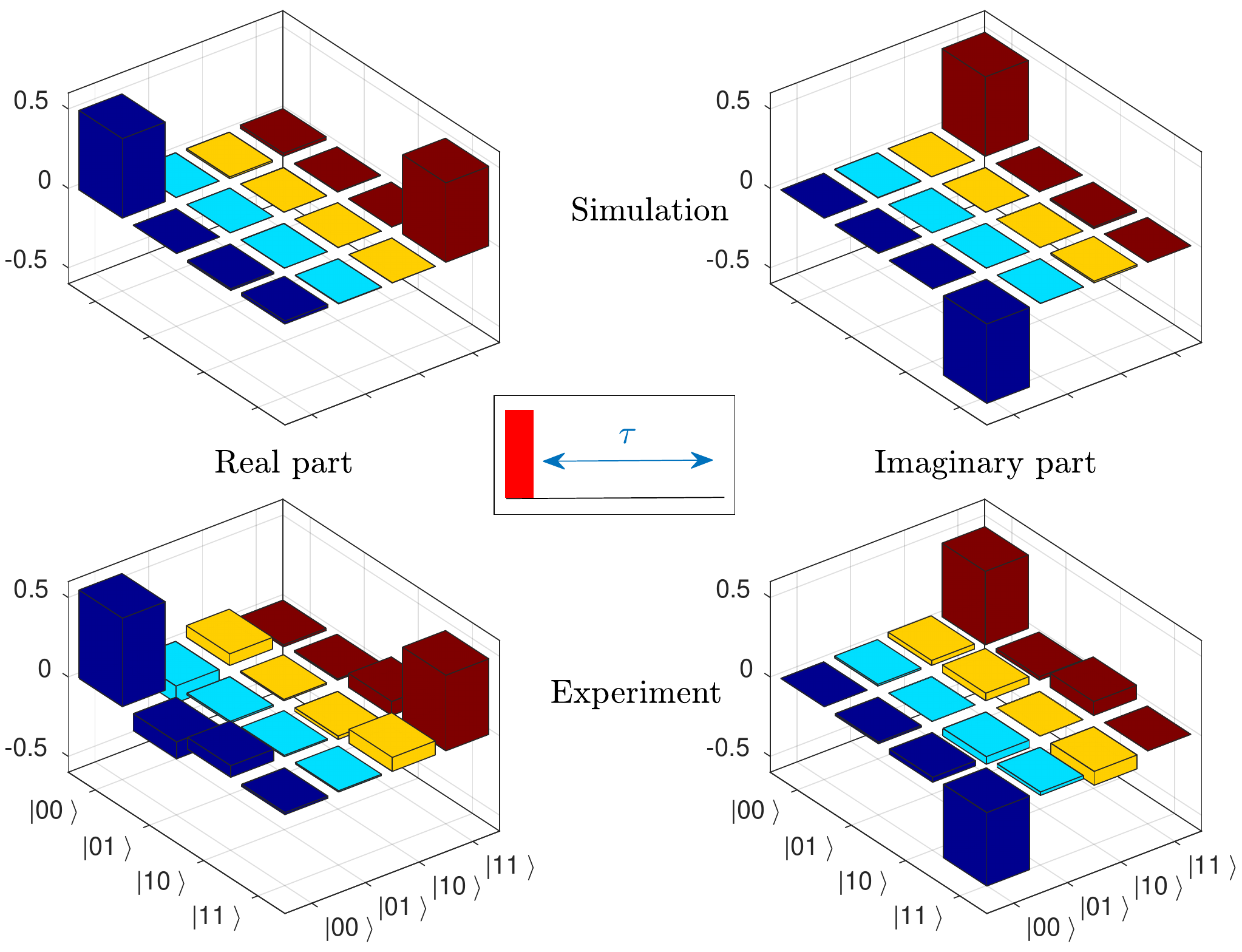}\caption{Simulated (top trace) and experimental (bottom trace) density matrices
of the Bell-type state $|\beta\rangle=(|00\rangle+i|11\rangle)/\sqrt{2},$
in the computational basis. The left column indicates the real part
of the density matrix and the right column indicates the imaginary
part. The pulse sequence to generate this state is shown in the inset:
the solid rectangle indicates a $90^{\circ}$ MW pulse on the electron
with a phase of $270^{\circ}$ and the delay corresponds to $U_{CR}(\pi)$.}
\label{bell} 
\end{figure}

\textit{Discussion.--}A typical quantum sensing or computing protocol
requires a combination of multiple gates. In such quantum circuits,
integration of our $U_{CR}(\alpha)$ with other gates that require
active control fields is a necessity. In the typical case where 2-qubit
gates are speed-limiting, efficient 2-qubit gates like the one introduced
here have a sigificant effect on the exectution time of the full protocol.
As an example, a Bell state can be prepared from an initial state
$\psi_{0}=|00\rangle$ by applying a Hadamard gate on the first qubit
followed by a CNOT gate targeting the second qubit \cite{nielsen,stolze}.
We here replace the Hadamard gate with the pseudo-Hadamard gate on
the electron spin and implement the CNOT gate by the free evolution.
The corresponding pulse sequence to prepare the Bell-type state is
shown in the inset of Fig. \ref{bell}. The solid rectangle is a $90^{\circ}$
MW pulse that creates an equal superposition between the electron
states $|0\rangle$ and $|1\rangle$. After a free evolution of duration
$\tau=4.545\,\mu$s, the ideal Bell-type state is $|\beta\rangle=(|00\rangle+i|11\rangle)/\sqrt{2}$.
The simulated density matrix $|\beta_{s}\rangle\langle\beta_{s}|$
obtained using the pulse sequence $(90^{\circ}-\tau$), where the
duration of the initial $90^{\circ}$ MW pulse is $0.125\,\mu$s,
is shown in the upper trace of Fig. \ref{bell}. Its state fidelity
with $|\beta\rangle\langle\beta|$ is 99.9\%. The reconstruction of
the experimental final state using full state tomography \cite{zhang2021fast}
is shown in the lower trace of Fig. \ref{bell} and its state fidelity
with $|\beta\rangle\langle\beta|$ is $96\%$. The experimental errors
are mainly due to the uncertainty of $\approx2\%$ due to the readout
gate operations required for quantum state tomography \cite{zhang2021fast}
and due to the photon counting statistics of $\approx3\%$.

For multi-qubit systems, the implementation of our fastest gates is
limited by the spectral resolution \cite{mizuochi2009coherence}.
In such cases, one can combine our gates with dynamical decoupling
to extend the $T_{2}^{*}$ \cite{zhang2014protected}; this will be
a subject of future work.

\textit{Conclusion.--}In conclusion, we have introduced a highly
efficient 2-qubit gate operation that uses only free evolution under
the static system Hamiltonian. Here, for a given Hamiltonian, we chose
computational basis states that are not the eigenstates of the system
Hamiltonian. Our scheme to implement the 2-qubit gate does not contribute
to the control overhead or to control errors. To demonstrate our scheme,
we choose a quantum register consisting of a single $^{13}$C spin
coupled to the electron spin of an NV center in diamond. The 2-qubit
gate efficiency derives from the hyperfine coupling that is much stronger
than the typical Rabi frequencies of the nuclear spins. Integration
of our individual controlled rotation gates with gates that require
active fields can improve the overall efficiency of the full quantum
computing or quantum sensing protocols.

\textit{Acknowledgments.--}This project has received funding from
the European Union's Horizon 2020 research and innovation programme
under grant agreement No 828946. The publication reflects the opinion
of the authors; the agency and the commission may not be held responsible
for the information contained in it.

SH and JZ contributed equally to this work.

 \bibliographystyle{apsrev}

\bibliography{ref_cr}

\end{document}